# Radiomics as a measure superior to the Dice similarity coefficient for tumor segmentation performance evaluation

Yoichi Watanabe[1] and Rukhsora Akramova[1]

[1]Department of Radiation Oncology, University of Minnesota Medical School, Minneapolis, MN, USA

**ABSTRACT**

*Background:* The segmentation accuracy of targets and healthy structures is one of the most essential requirements for high-quality radiotherapy delivery. The standard method for evaluation of the segmentation quality is to use similarity indices, and additional measures are desired to aid the selection of tools and improve the training of new physicians on this critical task.

*Purpose:* This study aims to propose Radiomics features as a superior measure for evaluating the segmentation ability of physicians and auto-segmentation tools compared to the commonly used Dice Similarity Coefficient (DSC).

*Methods:* First, the most reproducible radiomics features were selected as potential radiomics features for evaluating the segmentation accuracy by analyzing radiomics features calculated on two CT scans of ten lung tumors taken 15 minutes, which are available in the RIDER Data Library. Radiomics features were extracted using the PyRadiomics program. The selection was made by calculating the Concordance Correlation Coefficient (CCC). Next, we used the CT images of ten patients with nine segmentations created by different physicians or auto-segmentation tools for the segmentation performance study. DSC and the Intraclass Correlation Coefficient (ICC) of radiomics features were employed to compare and evaluate the segmentation similarity.

*Results:* The analysis revealed 206 radiomics features with a high CCC (or > 0.93) between two CT images, indicating robust reproducibility. Among these features, seven showed low ICC, suggesting an increased sensitivity to segmentation differences. The ICC of radiomics features exhibited greater sensitivity to segmentation changes compared to DSC. For example, the ICCs of the original shape sphericity, elongation, and flatness features ranged from 0.1177 to 0.991, 0.47 to 0.993, and 0.478 to 0.995, respectively. On the other hand, all DCS were larger than 0.778. The results indicated that the radiomics features could capture subtle variations in tumor segmentation characteristics, particularly in the features related to shape and energy, but DCS could not.

*Conclusion:* This study demonstrates the superiority of radiomics features with ICC as a measure for evaluating a physician's tumor segmentation ability and the performance of autosegmentation tools compared to the DSC. Radiomics features offer a more sensitive and comprehensive evaluation, providing valuable insights into tumor characteristics. Therefore, the new metrics can be used to evaluate new auto-segmentation methods and enhance trainees' segmentation skills in medical training and education.

# 1 INTRODUCTION

Many new autosegmentation methods using artificial intelligence technology are being developed. [1-3] To evaluate the performance of new tools, the need for a more sensitive and informative evaluation tool for tumor segmentation in medical imaging has become increasingly apparent.

It's essential to have a way to check the correctness and accuracy of segmentation. [4] The Dice similarity coefficient (DSC), the Surface Dice coefficient (SDC), and the Hausdorf distance (HD) are frequently employed to assess the similarity by quantifying the overlap between two segmentations. [5, 6] While DSC is a widely used metric for evaluating the similarity between auto-segmented and reference volumes, it has several limitations to consider when interpreting its results. [5, 7] The DSC treats all disagreements between the segmented volume and the reference volume equally, regardless of whether errors are systematic (consistent across multiple cases) or random (vary between cases). As a result, it cannot differentiate between these error types, potentially leading to misleading interpretations of the segmentation quality. The DSC primarily focuses on the overlap of segmented volumes and does not consider the clinical significance of these overlaps. It may not accurately reflect the practical utility of a segmentation algorithm in cases where minor discrepancies might have significant clinical implications, such as when segments are located near critical anatomical structures. The DSC does not provide information about the location of segmentation errors. A low DSC value could be due to errors occurring at the periphery of the segmented region, which might be less critical than errors near the center. The DSC can be sensitive to the choice of segmentation threshold and the resolution of the images. Variability in these factors can influence the calculated DSC values and make comparing different studies challenging. [5]

The abovementioned limitations highlight the need for a comprehensive evaluation beyond the DSC when assessing segmentation quality. We have identified radiomics features as a more comprehensive approach that provides valuable information about tumor size, shape, intensity, and texture characteristics. Considering the comprehensiveness of radiomics in evaluating image characteristics, [8] we hypothesize that radiomics features can offer a convenient means to compare and assess segmentations. Using the public data library RIDER [9], we demonstrate the superiority of radiomics features compared with DSC to evaluate the difference in segmentation.

# 2 METHODS AND MATERIALS
## 2.1 Data

This study used publicly available data kept in the RIDER (Reference Database to Evaluate Response) data library. [10] Only lung data were downloaded for the analysis. The data consisted of two datasets. The first set was two CT image data taken 15 minutes apart and one tumor segmentation made on the first CT image. The second set was nine segmentations made by three institutions using auto-segmentation tools. Each institution did tumor segmentation using institution-specific software by setting three different segmentation parameters. The two sets of data were prepared for ten patients. [9] Consequently, our data set contained ten pairs of CT image data sets and 90 segmentation data of lung tumors drawn on ten CT image sets. [11]

## 2.2 Radiomics

Radiomics utilizes mathematical formulas to characterize the shape, intensity, and texture with additional image filters, [12] as illustrated in Figure 1. This study employed 3D Slicer [13] for radiomics feature extraction and DSC calculations of the RIDER lung image data. The Radiomics module in 3D Slicer, SlicerRadiomics, is a specialized component that facilitates the extraction and analysis of radiomics features from medical images. It is a 3D Slicer implementation of PyRadiomics. [14] The module allows users to load medical image data (DICOM or other formats) and create or load segmentations representing regions of interest (ROIs) within the images. We calculated a range of radiomics features, including shape-based, intensity-based (or the first order), and texture-based metrics. CT images were resampled with 2 mm voxel size and filtered by Gaussian-of-Laplacian (LoG) 1,1,1 mm and wavelets. The Bin width was set as 25. Nine hundred forty-four radiomics features were obtained for each of the 110 image data sets (two for reproducibility and nine with nine different segmentations for ten patients).

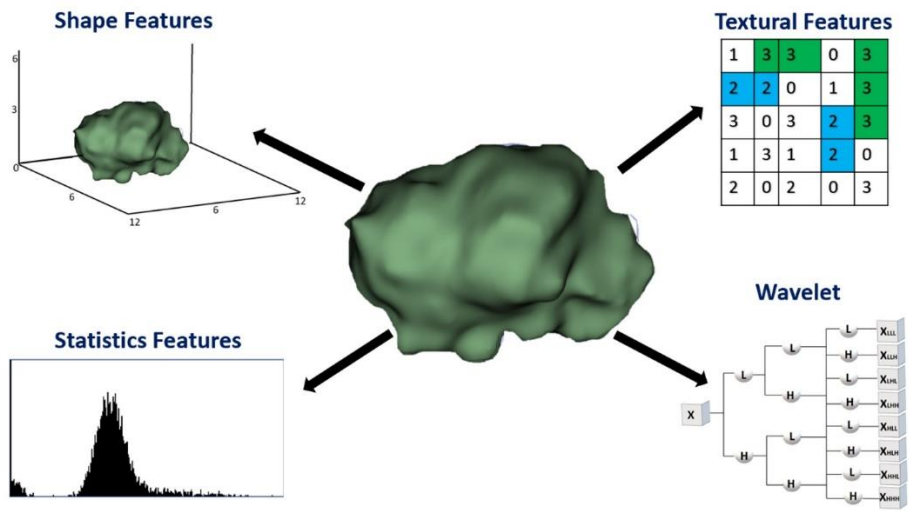

**Figure 1**   Radiomic features represent a tumor's intensity (statistics), shape, and texture, providing a comprehensive overview of its properties. CT images were resampled with 2 mm voxel size and filtered by LoG 1,1,1 mm and wavelets.

## 2.3 Selection of radiomics features

Figure 2 shows two CT images taken 15 minutes apart for one patient. The tumor was segmented independently for CT #1 and CT #2. Because the position and shape of the tumor are different between the two CT images, the radiomics feature values calculated on these two CT images can be different even for the same tumor contour. Therefore, first, we selected highly repeatable and reproducible radiomics features by computing the Concordance Correlation Coefficient (CCC) or Correlation Index (CI) of ten CT pairs for all 944 radiomics features. [15, 16] The radiomics features with CCC values higher than 0.93 were saved for further analysis.

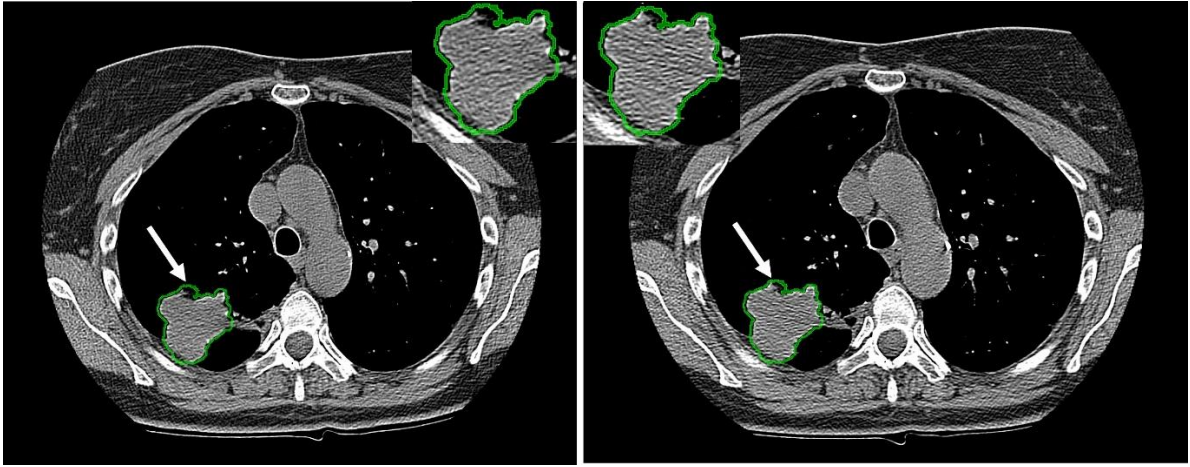

**Figure 2**     Two image sets of the same patient used for CCC calculations. CT data on the left and right were taken 15 minutes apart. White arrows point to areas of the tumor that are different.

## 2.4 Calculations of ICC

Figure 3 shows two segmentations drawn for the same CT image data. We can easily recognize slight but non-negligible differences in the tumor's shape. To quantify the difference between two segmentations drawn on the same image, we calculated DSC and radiomics feature values. DSC was calculated for a pair of nine segmentations using the 3D Slicer, resulting in 36 values per patient. The calculations were repeated for all ten patients. Hence, there were 36x10 values of DSC.

First, we obtained nine values per radiomics feature per patient for the first CT data and nine lung tumor segmentation data. There was a 9 x 10 matrix of values for ten patients, which provided one intraclass correlation coefficient (ICC) per radiomics feature. [17] Next, we calculated the ICC for 36 pairs of nine segmentation, resulting in 36 ICC values per radiomics feature. [18]

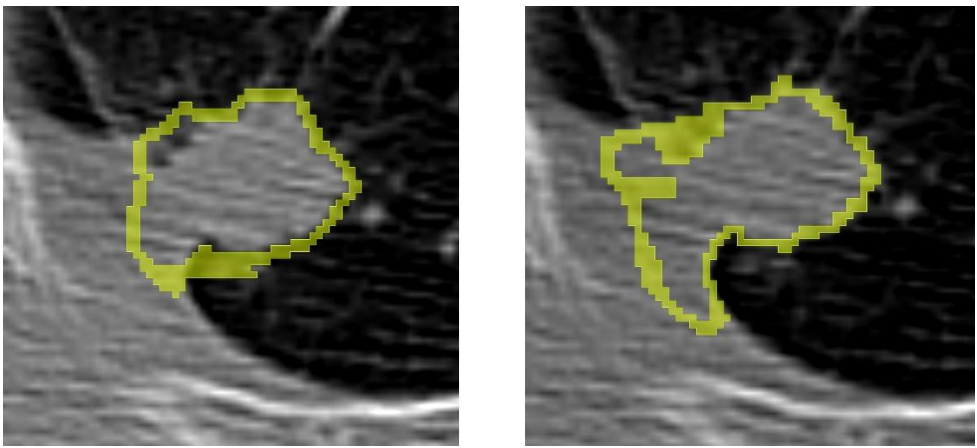

**Figure 3**     Two contours on the same CT image set of one patient are different.

## 2.5 Statistical analysis

All statistical analyses were done using standard Python library routines and the ICC package available at GitHub. [19]

## 3 RESULTS

Our analysis of the CCC for ten sets of CT image data taken 15 minutes apart revealed 206 radiomics features with CCC greater than 0.93. These 206 features are considered to have good reproducibility.

To compare nine segmentation methods applied to the CT data of ten patients, we selected seven radiomics features with the lowest ICC minimum, the minimum ICC among 36 pairs of segmentation data. The low ICC indicates low reproducibility of feature values for different segmentations. Table 1 lists the seven radiomics features selected for the analysis.

| Image type | Feature Class | Feature Name | CCC | ICC | Lower Bound | Upper Bound | Min ICC |
|---|---|---|---|---|---|---|---|
| original | shape | Elongation | 0.979 | 0.719 | 0.551 | 0.879 | 0.475 |
| original | shape | Flatness | 0.985 | 0.757 | 0.600 | 0.898 | 0.478 |
| original | shape | Sphericity | 0.935 | 0.600 | 0.392 | 0.815 | 0.177 |
| wavelet-LLH | firstorder | TotalEnergy | 0.963 | 0.821 | 0.691 | 0.928 | 0.673 |
| wavelet-HLL | firstorder | TotalEnergy | 0.950 | 0.789 | 0.645 | 0.913 | 0.592 |
| wavelet-HLH | firstorder | TotalEnergy | 0.965 | 0.771 | 0.620 | 0.905 | 0.635 |
| wavelet-LLL | gldm | Small Dependence Low Gray Level Emphasis | 0.942 | 0.874 | 0.774 | 0.951 | 0.675 |

**Table 1** CCC, ICC, Lower and Upper Bound of seven selected radiomics features with 95% Confidence interval. Min ICC is the smallest among the ICCs for 36 pairs of 9 segmentations.

There were 36 ICC values per radiomics feature. One DSC value was obtained for each pair of segmentation per patient; hence, there were 36 x 10 DSC values for 10 patients. For comparison with the radiomics method, we took an average of 10 patients to have 36 DSC values. In Figure 4, we plotted 36 ICC and DSC for 36 pairs of segmentations. The ICC of radiomics features exhibited greater sensitivity to segmentation changes than the Dice coefficient as the ICC values are wildly spread out, i.e., from 0.0 to 1.0. In contrast, all DSC values were in a narrow range of 0.75 to 1.0. According to the standard system, all the segmentations are in good or very good agreement with DSC > 0.7 [20], whereas the ICC indicated that some segmentations were very different from others. Notably, the agreement is poor with ICC < 0.5 for Original shape sphericity, Original shape elongation, and Original shape flatness, or moderate with ICC < 0.75 for Wavelet LHH/HLL first order Total Energy. [21] For example, the ICCs of the original shape sphericity, elongation, and flatness features ranged from 0.1177 to 0.991, 0.47 to 0.993, and 0.478 to 0.995, respectively.

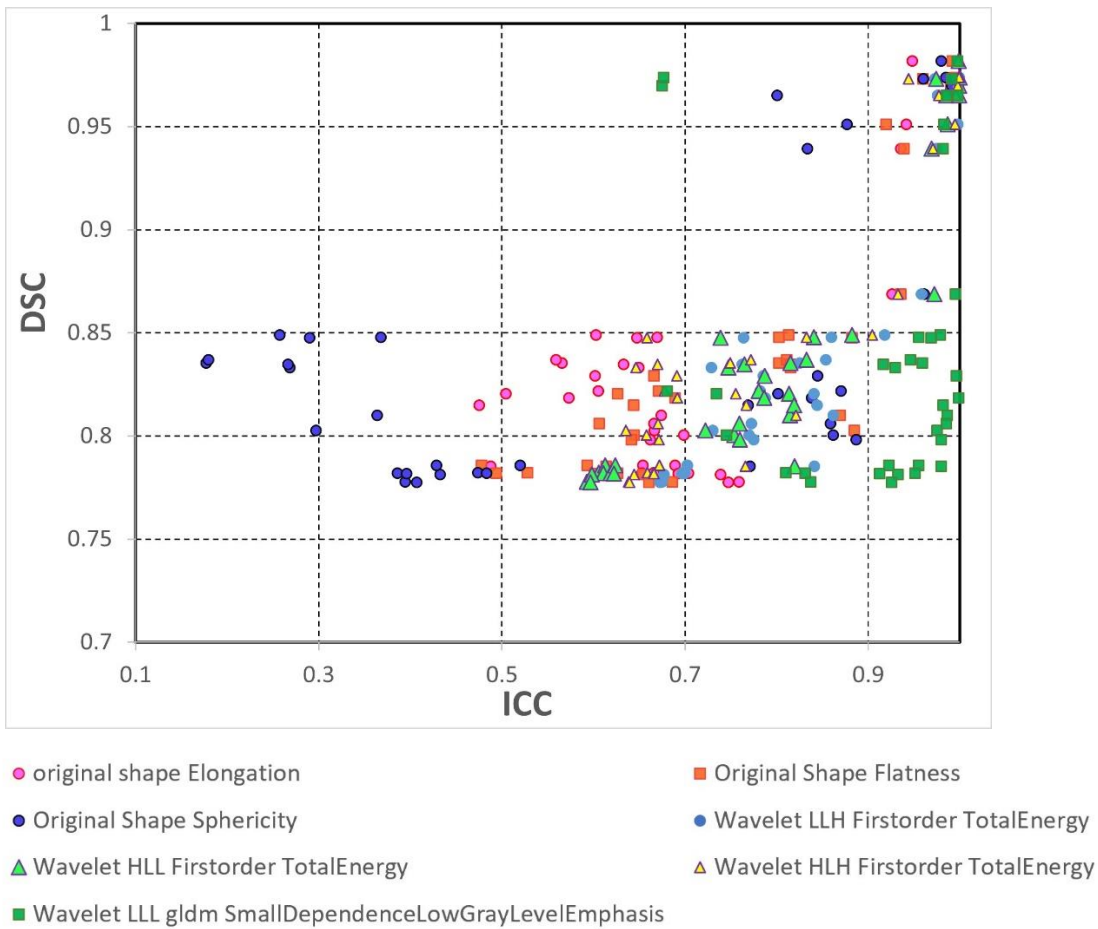

**Figure 4**  ICC of seven radiomics features vs DSC. There are 36 ICC values per feature and 36 DSCs independent of the radiomics features.

To further demonstrate the high sensitivity of radiomics features to differences in the segmentation, we plotted heatmaps of three features (Shape original sphericity, elongation, and total energy) in Figure 5. The figure indicated the degree of correlation between 36 pairs of 9 segmentations. Note that an average feature value of 10 patients was used for the correlation analysis. The figure showed three distinguishable classes among nine segmentations. Class A: segmentations 1 to 3, Class B: 4 to 6, and Class C: 7 to 9 belong to separate classes/groups. It is clear, especially in Figure 5 (b): Elongation and (c) Total energy plots. The results suggest that different software and persons segmented three classes of segmentations. In other words, we can conclude that three groups created the nine segmentations, as the similarity of the values for the group of the first three segmentations and the second and third groups is noticeable.

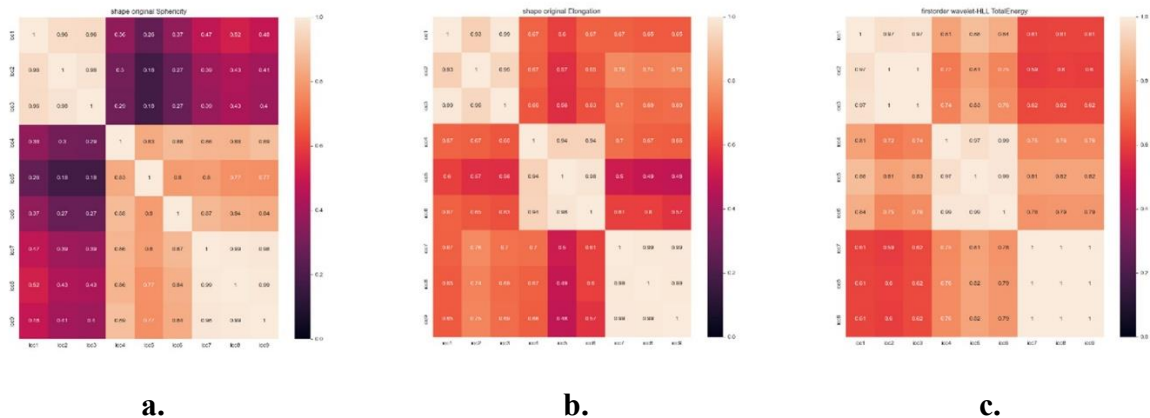

                **a.**                               **b.**                               **c.**

**Figure 5**    Heat maps of Radiomics features: (a) Shape original sphericity, (b) Shape Original Elongation, (c) FirstOrder Wavelet HLL total energy

## 4    EXAMPLE APPLICATION OF RADIOMICS

In the realm of medical imaging, the assessment of tumor segmentation quality is of paramount importance. This study has convincingly demonstrated the efficacy of Radiomics features in offering a meticulous evaluation of tumor segmentations. However, it is imperative to consider alternative approaches that complement this assessment to provide a well-rounded evaluation. The Intraclass Correlation Coefficient (ICC) is a valuable tool for measuring the agreement and consistency between different sets of segmentations. Within our context, ICC can be viewed as a benchmark for segmentation accuracy.

Imagine a scenario where ICC values between segmentation pairs, such as 1-2 or 2-3, exhibit a high level of agreement. These pairs can be deemed as reference or "Gold Standard" segmentations, signifying their accuracy and reliability in segmentation. Consequently, other segmentation methods can be evaluated by comparing their ICC values to those of the reference pairs. Segmentation techniques that closely align with these reference ICC values can be regarded as more accurate and consistent in capturing the intricacies of tumor characteristics.

For example, let us examine ICC values from the current analyses in Table 2. We arbitrarily set segmentation #1 as the gold standard. From the columns icc1-7, 1-8, and 1-9, all ICC values except wavelet-LLL gldm are less than 0.75, with some less than 0.5. It implies the segmentation method used for segmentations 7, 8, and 9 is poor quality compared to the gold standard. The method could be a physician or software. Hence, with ICC, it is easy to evaluate the performance of methods/or person who does segmentation. If we used DSC instead, these poor-performing methods could be considered a good segmentation tool.

| Image type | Feature Class | Feature Name | icc1-2 | icc1-3 | icc1-4 | icc1-5 | icc1-6 | icc1-7 | icc1-8 | icc1-9 |
|---|---|---|---|---|---|---|---|---|---|---|
| *original* | *shape* | *Elongation* | 0.925 | 0.991 | 0.673 | 0.602 | 0.669 | 0.666 | 0.653 | 0.652 |
| *original* | *shape* | *Flatness* | 0.935 | 0.959 | 0.869 | 0.812 | 0.802 | 0.527 | 0.477 | 0.494 |
| *original* | *shape* | *Sphericity* | 0.961 | 0.95 | 0.363 | 0.257 | 0.367 | 0.473 | 0.520 | 0.483 |
| *wavelet-LLH* | *firstorder* | *TotalEnergy* | 0.958 | 0.972 | 0.861 | 0.918 | 0.859 | 0.697 | 0.702 | 0.696 |
| *wavelet-HLL* | *firstorder* | *TotalEnergy* | 0.972 | 0.974 | 0.814 | 0.882 | 0.840 | 0.605 | 0.612 | 0.610 |
| *wavelet-HLH* | *firstorder* | *TotalEnergy* | 0.932 | 0.944 | 0.820 | 0.904 | 0.832 | 0.659 | 0.672 | 0.665 |
| *wavelet-LLL* | *gldm* | *SmallDependenceLowGrayLevelEmphasis* | 0.995 | 0.991 | 0.986 | 0.978 | 0.954 | 0.809 | 0.954 | 0.950 |
| | | *Average* | 0.954 | 0.967 | 0.788 | 0.805 | 0.785 | 0.639 | 0.657 | 0.652 |

**Table 2** Interclass correlation (ICC) for eight pairs of segmentations of seven selected radiomics

## 5 DISCUSSION

### 5.1 DSC and radiomics features as similarity metrics

This study employed two similarity metrics, the DSC and Radiomics features, to evaluate and compare tumor segmentations in medical images. The DSC, widely used in medical image segmentation tasks, measures the overlap between two segmentations by quantifying the similarity between the segmented region and the ground truth mask. It provides a value ranging from 0 to 1, where 1 indicates a perfect overlap, and 0 represents no overlap between the segmentations.

Radiomics features, on the other hand, are quantitative and high-dimensional features extracted from medical images using advanced image processing techniques. These features capture various aspects of the tumor's characteristics, including size, shape, intensity, and texture, providing valuable information for assessing tumor properties beyond simple overlap measurements. The Radiomics features are derived from the specific segmentation area of the tumor, utilizing the region of interest defined by the physician or radiologist during segmentation. As a result, they offer a more comprehensive and detailed representation of the tumor's characteristics, contributing to a deeper understanding of the segmentation quality.

Figure 6 presents two segmentations overlayed on the same CT image data (segmentation #1 in yellow and segmentation #5 in green shaded areas). Different specialists or segmentation software created these two segmentations and did not entirely match. While the DSC indicates an 88% overlap, which is unreasonably high in this picture, the Radiomics feature's value significantly differs for these segmentations. For example, Sphericity features, which describe the roundness or compactness of a tumor and are calculated as the ratio between the surface area of a sphere with the same volume as the tumor and the actual surface area, Wavelets first-order energy Radiomics features, which provide information about the energy of the pixel intensity values in a specific frequency range, were found to be especially sensitive to segmentation changes. Table 3 lists the radiomics feature values of the two segmentations for this example.

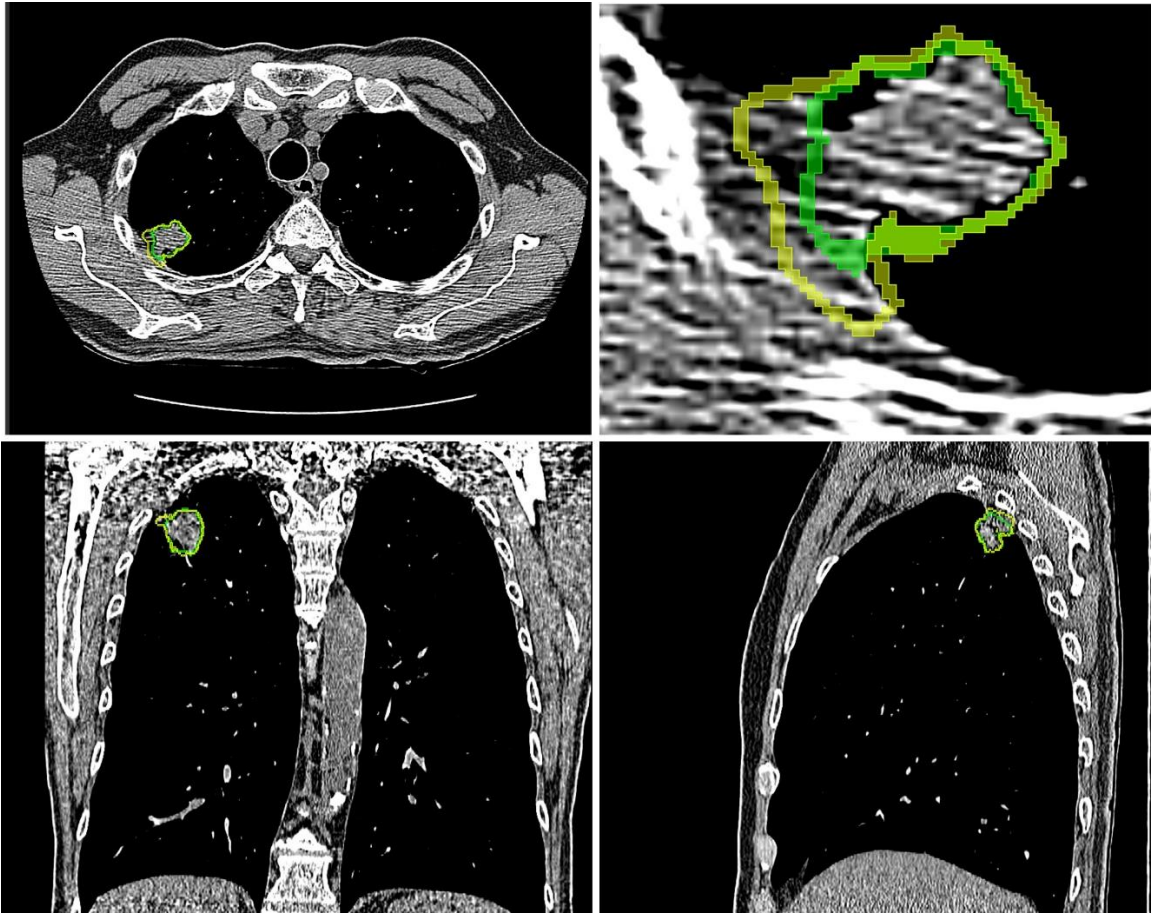

**Figure 6**      Example of two segmentations of the same tumor.

| Image type | Feature Class | Feature Name | Segmentation 1 | Segmentation 5 |
|---|---|---|---|---|
| original | shape | Elongation | 0.8445731 | 0.9180314 |
| original | shape | Flatness | 0.6902414 | 0.7137111 |
| original | shape | Sphericity | 0.7076588 | 0.6624693 |
| wavelet-LLH | firstorder | TotalEnergy | 57595766 | 76010437 |
| wavelet-HLL | firstorder | TotalEnergy | 173615173 | 263878340 |
| wavelet-HLH | firstorder | TotalEnergy | 26963893 | 35211480 |
| wavelet-LLL | gldm | Small Dependence Low Gray Level Emphasis | 0.0003348 | 0.0002372 |

**Table 3**   Seven radiomics values of two segmentations for the example case.

**5.2 Radiomics as an evaluation tool of segmentation ability and skills**

The implications of this study extend to the realm of medical education. As budding physicians, radiologists, and healthcare professionals undergo training, acquiring accurate and precise tumor segmentation skills becomes paramount. Incorporating Radiomics features into the evaluation process, juxtaposed against the widely used DSC, can enable educators to develop comprehensive assessment tools. Medical training programs can employ this novel approach to give trainees constructive feedback on their tumor segmentation proficiency, thereby identifying areas of improvement and tailoring personalized learning plans to enhance their segmentation skills.

Beyond education, this research bears practical implications for clinical practice and quality assurance in segmentation software employing various algorithms. Healthcare facilities that utilize automated or semi-automated segmentation software can leverage Radiomics features as an evaluation metric when comparing algorithm-generated segmentations with human-generated ones. This approach facilitates the identification of algorithm strengths and weaknesses, guiding further optimization and refinement. Moreover, healthcare institutions can employ this methodology to conduct routine audits, ensuring the accuracy and consistency of segmentation results across different software platforms and imaging techniques.

**5.3 Limitations and future studies**

There are several limitations to the current study. First, we used only DSC as a similarity index, but other indices are proposed for contour similarity tests, such as Houdorf distance and Surface Dice similarity coefficient [6, 22]. Comparison of other indices with the radiomics-based evaluation will be studied. Secondly, our application was limited to small lung tumors with CT. Hence, a study is needed for tumors with a larger volume, in different anatomical sites, and imaging techniques other than CT. Lastly, we plan to use the new evaluation tool to improve the segmentation skills of radiation oncology residents and other professionals.

# 6 CONCLUSIONS

The findings of this study demonstrate the superiority of Radiomics features as an evaluation tool for different tumor segmentation methods in medical imaging when compared to the Dice similarity coefficient. Radiomics features provide a more sensitive and informative approach, capturing intricate details of tumor characteristics such as size, shape, texture, and intensity. Radiomics features better detect and capture subtle variations or differences in these tumor properties than the Dice similarity coefficient.

The practical application of this research includes evaluating segmentation abilities during medical training and education and identifying weaknesses in segmentation programs that employ various algorithms. Future work should explore alternative Dice similarity coefficients based on average and maximum distance, additional path length, and changes in surface, and the extension of this methodology to other anatomical locations and imaging modalities beyond lung CT imaging.

In summary, this study not only underscores the superiority of Radiomics features in segmentation evaluation but also emphasizes the significance of considering ICC indices as a complementary approach, thereby contributing to advancing tumor segmentation assessment in medical imaging.

## ACKNOWLEDGMENT

None

## CONFLICT OF INTEREST STATEMENT

The authors declare no conflict of interest.

## REFERENCES

1. Sharp, G., et al., *Vision 20/20: perspectives on automated image segmentation for radiotherapy.* Med Phys, 2014. **41**(5): p. 050902.
2. Wong Yuzhen, N. and S. Barrett, *A review of automatic lung tumour segmentation in the era of 4DCT.* Rep Pract Oncol Radiother, 2019. **24**(2): p. 208-220.
3. Harrison, K., et al., *Machine Learning for Auto-Segmentation in Radiotherapy Planning.* Clin Oncol (R Coll Radiol), 2022. **34**(2): p. 74-88.
4. Brown, K.H., et al., *Assessment of Variabilities in Lung-Contouring Methods on CBCT Preclinical Radiomics Outputs.* Cancers (Basel), 2023. **15**(10): p. 2677.
5. Sherer, M.V., et al., *Metrics to evaluate the performance of auto-segmentation for radiation treatment planning: A critical review.* Radiother Oncol, 2021. **160**: p. 185-191.
6. Mackay, K., et al., *A Review of the Metrics Used to Assess Auto-Contouring Systems in Radiotherapy.* Clinical Oncology, 2023. **35**(6): p. 354-369.
7. Taha, A.A. and A. Hanbury, *Metrics for evaluating 3D medical image segmentation: analysis, selection, and tool.* BMC medical imaging, 2015. **15**: p. 29-29.

8.  Lambin, P., et al., *Radiomics: extracting more information from medical images using advanced feature analysis.* Eur J Cancer, 2012. **48**(4): p. 441-6.
9.  Armato, S.G., 3rd, et al., *The Reference Image Database to Evaluate Response to therapy in lung cancer (RIDER) project: a resource for the development of change-analysis software.* Clin Pharmacol Ther, 2008. **84**(4): p. 448-56.
10. Nolan, T. *RIDER Collections*. 2023; May 24, 2023:[Available from: https://wiki.cancerimagingarchive.net/display/public/rider+collections.
11. Kirby, J., *Segmentation data in RIDER lung CT database*. 2023, Imaging Data Commons.
12. Aerts, H.J., et al., *Decoding tumour phenotype by noninvasive imaging using a quantitative radiomics approach.* Nat Commun, 2014. **5**: p. 4006.
13. Nikolov, S., et al., *Clinically Applicable Segmentation of Head and Neck Anatomy for Radiotherapy: Deep Learning Algorithm Development and Validation Study.* J Med Internet Res, 2021. **23**(7): p. e26151.
14. van Griethuysen, J.J.M., et al., *Computational Radiomics System to Decode the Radiographic Phenotype.* Cancer Res, 2017. **77**(21): p. e104-e107.
15. Zhao, B., et al., *Evaluating variability in tumor measurements from same-day repeat CT scans of patients with non-small cell lung cancer.* Radiology, 2009. **252**(1): p. 263-72.
16. Balagurunathan, Y., et al., *Test-retest reproducibility analysis of lung CT image features.* J Digit Imaging, 2014. **27**(6): p. 805-23.
17. Owens, C.A., et al., *Lung tumor segmentation methods: Impact on the uncertainty of radiomics features for non-small cell lung cancer.* PLOS ONE, 2018. **13**(10): p. e0205003.
18. Liljequist, D., B. Elfving, and K. Skavberg Roaldsen, *Intraclass correlation - A discussion and demonstration of basic features.* PLoS One, 2019. **14**(7): p. e0219854.
19. Dafflon, J. and W. Hugo Lopez Pinaya. *Mind-the-Pineapple/ICC* 2023; Available from: https://github.com/Mind-the-Pineapple/ICC.
20. Bao, D., et al., *Baseline MRI-based radiomics model assisted predicting disease progression in nasopharyngeal carcinoma patients with complete response after treatment.* Cancer Imaging, 2022. **22**(1): p. 10.
21. Koo, T.K. and M.Y. Li, *A Guideline of Selecting and Reporting Intraclass Correlation Coefficients for Reliability Research.* J Chiropr Med, 2016. **15**(2): p. 155-63.
22. Kiser, K.J., et al., *Novel Autosegmentation Spatial Similarity Metrics Capture the Time Required to Correct Segmentations Better Than Traditional Metrics in a Thoracic Cavity Segmentation Workflow.* J Digit Imaging, 2021. **34**(3): p. 541-553.